
\input harvmac
\input epsf

\noblackbox
\def\IZ{\relax\ifmmode\mathchoice
{\hbox{\cmss Z\kern-.4em Z}}{\hbox{\cmss Z\kern-.4em Z}}
{\lower.9pt\hbox{\cmsss Z\kern-.4em Z}}
{\lower1.2pt\hbox{\cmsss Z\kern-.4em Z}}\else{\cmss Z\kern-.4em Z}\fi}
\def\IC{\relax{\rm l\kern-.42em C}}
\def\IP{\relax{\rm I\kern-.18em P}}
\def\IR{\relax{\rm I\kern-.18em R}}
\font\cmss=cmss10 \font\cmsss=cmss10 at 7pt

\def\frac#1#2{{#1 \over #2}}

\def\ee{{E_8 \times E_8}}
\def\aE{{{\tilde L}E_8}}
\def\LE{{LE_8}}

\lref\WittenMD{
E.~Witten,
``On flux quantization in $M$-Theory and the effective action,''
J.\ Geom.\ Phys.\  {\bf 22}, 1 (1997)
arXiv:hep-th/9609122.
}

\lref\WittenVG{
E.~Witten,
``Duality relations among topological effects in string theory,''
JHEP {\bf 0005}, 031 (2000)
arXiv:hep-th/9912086.
}

\lref\DiaMoWitten{
D.~E.~Diaconescu, G.~W.~Moore and E.~Witten,
``E(8) gauge theory, and a derivation of $K$-Theory from $M$-Theory,''
arXiv:hep-th/0005090
}

\lref\DiaconescuWZ{
D.~E.~Diaconescu, G.~W.~Moore and E.~Witten,
``A derivation of $K$-Theory from $M$-Theory,''
arXiv:hep-th/0005091.
}

\lref\WittenCD{
E.~Witten,
``D-branes and $K$-Theory,''
JHEP {\bf 9812}, 019 (1998)
arXiv:hep-th/9810188.
}

\lref\MooreGB{
G.~W.~Moore and E.~Witten,
``Self-duality, Ramond-Ramond fields, and $K$-Theory,''
JHEP {\bf 0005}, 032 (2000)
arXiv:hep-th/9912279.
}

\lref\FabingerJD{
M.~Fabinger and P.~Ho{\v r}ava,
``Casimir effect between world-branes in Heterotic $M$-theory,''
Nucl.\ Phys.\ B {\bf 580}, 243 (2000)
arXiv:hep-th/0002073.
}

\lref\MadridGuys{
C.~Gomez and J.~J.~Manjarin,
``Dyons, $K$-Theory and $M$-Theory''
arXiv:hep-th/0111169
}

\lref\MMS{
J.~Maldacena, G.~W.~Moore and N.~Seiberg,
``Geometrical interpretation of D-branes in gauged WZW models,''
JHEP {\bf 0107}, 046 (2001)
arXiv:hep-th/0105038.
}

\lref\BoMathai{
P.~Bouwknegt and V.~Mathai,
``D-branes, B-fields and twisted $K$-Theory,''
JHEP {\bf 0003}, 007 (2000)
arXiv:hep-th/0002023.
}

\lref\SquareThree{
J.~Evslin and U.~Varadarajan,
``$K$-Theory and S-duality: Starting over from square 3,''
arXiv:hep-th/0112084.
}

\lref\HW{
P.~Ho{\v r}ava and E.~Witten,
``Eleven-Dimensional Supergravity on a Manifold with Boundary,''
Nucl.\ Phys.\ B {\bf 475}, 94 (1996)
arXiv:hep-th/9603142.
}

\lref\HoravaQA{
P.~Ho{\v r}ava and E.~Witten,
``Heterotic and type I string dynamics from eleven dimensions,''
Nucl.\ Phys.\ B {\bf 460}, 506 (1996)
arXiv:hep-th/9510209.
}

\lref\WittenTools{
E.~Witten,
``Topological Tools In Ten-Dimensional Physics,''
Int.\ J.\ Mod.\ Phys.\ A {\bf 1}, 39 (1986).
}

\lref\Rosenberg{
J.~Rosenberg, 
{\it Continuous trace algebras from the bundles theoretic point of view},
Jour.~Austr.~Math.~Soc.,~{\bf 47} (1989), 368-381
}

\lref\FreedWitten{
D.~S.~Freed and E.~Witten,
``Anomalies in string theory with D-branes,''
arXiv:hep-th/9907189.
}

\lref\MeWIP{
A.~Adams,
{\it Work In Progress}
}

\lref\HoravaUday{
P.~Ho{\v r}ava and U.~Varadarajan,
{\it Work in Progress}
}

\lref\Petrotic{
A.~Adams, M.~Fabinger, P.~Horava and U.~Varadarajan,
{\it Work in Progress}
}

\lref\HoravaEight{
P.~Ho{\v r}ava,
``$M$-Theory as a holographic field theory,''
Phys.\ Rev.\ D {\bf 59}, 046004 (1999)
arXiv:hep-th/9712130.
}

\lref\HoravaJY{
P.~Horava,
``Type IIA D-branes, $K$-Theory, and matrix theory,''
Adv.\ Theor.\ Math.\ Phys.\  {\bf 2}, 1373 (1999)
arXiv:hep-th/9812135.
}

\lref\LoopBook{
A.~Pressley and G.~Segal,
``Loop Groups,''
{\it Oxford, UK: Clarendon (1988) 318 p. (Oxford Mathematical Monographs).}
}

\lref\GutperleSX{
M.~Gutperle and A.~Strominger,
``Fluxbranes in string theory,''
JHEP {\bf 0106}, 035 (2001)
arXiv:hep-th/0104136.
}

\lref\WittenFiveBrane{
E.~Witten,
``Five-brane effective action in $M$-Theory,''
J.\ Geom.\ Phys.\  {\bf 22}, 103 (1997)
arXiv:hep-th/9610234.
}

\lref\TaylorZA{
W.~Taylor,
``D2-branes in B fields,''
JHEP {\bf 0007}, 039 (2000)
arXiv:hep-th/0004141.
}

\Title{\vbox{\baselineskip12pt
\hbox{hep-th/0203218}
\hbox{NSF-ITP-02-04}
\hbox{UCB-PTH-02-05}
\hbox{SLAC-PUB-9147}
}}{\vbox{
\centerline{The Loop Group of $E_8$ and $K$-Theory from $11d$}
}}
\centerline{Allan Adams\foot{allan@slac.stanford.edu}~ and~ Jarah Evslin\foot{jarah@uclink.berkeley.edu}}
\medskip
\centerline{$^1$~Dept. of Physics and SLAC, Stanford University, Stanford CA 94305/94309}
\centerline{$^2$~Dept. of Math, UC Berkeley, and Theory Group, LBNL, Berkeley CA 94720}
\bigskip
\bigskip
\bigskip

\noindent
We examine the conjecture that an $11d$ $E_8$ bundle, appearing in the calculation of 
phases in the $M$-Theory partition function, plays a physical role in $M$-Theory, focusing on 
consequences for the classification of string theory solitons.
This leads for example to a classification of IIA solitons in terms of that of $\LE$ bundles in $10d$. 
Since $K(\IZ,2)$ approximates $\LE$ up to $\pi_{14}$, this reproduces the $K$-Theoretic 
classification of IIA D-branes while treating NSNS and RR solitons more symmetrically
and providing a natural interpretation of $G_0$ as the central extension of $\aE$.

\Date{March 2002}

\newsec{Introduction and Motivation}

In this note we study the classification of solitons in string theory
and $M$-Theory.  Our starting point is the intersection of two suggestive results.
First, as argued by Witten \WittenMD\WittenVG\ and more extensively by Diaconescu, 
Moore and Witten \DiaMoWitten\DiaconescuWZ, certain subtle phases in the $M$-Theory 
partition function
suggest a 
connection to an $E_8$ gauge theory over a $12d$ manifold $Z$ bounded
by $Y$. This follows from the fact that $E_8$ bundles in $12d$ 
are specified topologically by their Chern-Simons $3$-form \WittenTools, 
so that the calculation of these $M$-Theory phases as
sums over topologically distinct $M$-Theory $3$-form configurations takes a 
natural form in terms of the index theory of $12d$ $E_8$ bundles.
That this $E_8$ index theory result agreed precisely with a very different 
calculation based on IIA $K$-Theory led Diaconescu, Moore and Witten to suggest 
a deeper connection between the $M$-Theory $3$-form and the Chern-Simons $3$-form 
of a $12d$ $E_8$ bundle.  
Since the index calculation depends only on $\partial Z=Y$, the physical
data lies in the restriction of the $12d$ bundle to an $E_8$ bundle in $11d$.

Secondly, it is commonly believed that the $K$-Theory of $\IC P^\infty \sim K(\IZ,2)$ bundles
classifies D-Brane configurations in Type IIA string theory, as argued in \WittenCD\MooreGB\ and 
phrased in terms of $K(\IZ,2)$ in \BoMathai.
However, the physical connection of the group $K(\IZ,2)$ to $M$-Theory is 
unclear.  Moreover, as fleshed out in a beautiful paper by
Maldacena, Moore and Seiberg \MMS, the Atiyah-Hirzebruch Spectral Sequence 
(AHSS) construction of the $K$-Theoretic classification 
of Type II RR solitons involves anomaly cancellation conditions in an intimate and beautiful way.  
How this relates to the proposal of \BoMathai\ is again unclear.

These lines of reasoning beg to be connected. As a first hint, note that $K(\IZ,2)$ 
and $\LE$ are homotopically identical up to
$\pi_{14}$.~\foot{$\LE$ denotes the loop group of $E_8$, and $\aE$ its centrally extended 
generalization. We describe their low-dimensional topology below; 
for a complete discussion, see eg \LoopBook.},\foot{We are deeply indebted to Petr Ho{\v r}ava 
for insightful discussions during early stages of this work suggesting looking at
the loop group of $E_8$ as an $M$-Theoretic alternative to the stringy picture of 
$K(\IZ,2)$ arising from an infinite number of 
unstable D$9$-branes \HoravaJY.  For a discussion of possible relations 
between these two pictures and their implications for supersymmetry and $11d$ dynamics, 
see \HoravaUday.}  Thus the classification of $\LE$ bundles over $10$-manifolds 
agrees with that of $\IC P^\infty$ bundles.  Further, up to important questions of 
central extension and torsion which we address below, the classification of $\LE$ bundles
over $10$-manifolds is precisely the classification of $E_8$ 
bundles over $11$-manifolds with a compatible circle action.
Thus the classification of solitons and the cancellation of anomalies in
$M$-Theory and IIA (and Heterotic, as we shall see),
as well as the relationship between these as revealed
by the AHSS, can all be phrased in terms of a single $E_8$ structure in $11d$.  
That an $11d$ $E_8$ bundle ties together so many pieces of the $M$-Theory puzzle 
strongly supports the conjecture that an $11d$ $E_8$ bundle plays a physical role 
in $M$-Theory, and should be reflected in its fundamental degrees of freedom.

Taking this seriously thus leads us to conjecture that 
the classification of RR and NSNS solitons in IIA derives from the classification 
of $\LE$ bundles over $10$-manifolds. 
This generalizes the accepted $K$-Theoretic classification 
of RR solitons (and adds to growing evidence that $K$-Theory at least approximately 
respects IIB S-duality, suggesting that $K$-Theory plays some role even beyond weak coupling)
while leading to novel predictions about the complete classification of IIA solitons,
including the interpretation of the cosmological ``constant''\foot{Since the dilaton 
is not constant in the presence of $D8$-branes, this
should properly be called a cosmological term rather than a cosmological constant.}
$G_0$ of (massive) IIA as the central charge of $\aE$,
and several constraints relating  torsion in $M$-Theory, $\aE$ and IIA.

In the remainder of this note we present further motivation for these conjectures
and show how such a framework reproduces and extends the familiar classification of solitons in 
$M$-Theory and its $10d$ descendants\foot{For earlier thoughts on the role of $E_8$ 
in $M$-Theory, see eg \MadridGuys\HoravaEight\FabingerJD.  See also \HoravaUday\Petrotic
for related current work}. Of course, $11d$ SUSY does not to play well with gauge bundles, and 
it is difficult to see how a dynamical bundle can coexist with $32$ supercharges.
(For further thoughts along these lines see eg \HoravaEight\HoravaUday.)
However, objects to which the $E_8$ gauge connection couples 
in $M$-Theory and the string theory generically violate at least half of the supercharges, 
so we might expect to see gauge bundle information only in situations with reduced supersymmetry. 
In any case, the resolution is unclear, so we
restrict ourselves in the following to studying the soliton classification,
leaving questions of dynamics and SUSY to future work.
We begin by reviewing the topological classification of 
$E_8$ bundles over $11$-manifolds.

\newsec{The Topological Classification of $E_8$ Bundles in $11d$}

$E_8$ has exceptionally simple low-dimensional topology.  In particular, its only 
non-trivial homotopy group below dimension $15$ is $\pi_3(E_8) = \IZ$.  The basic
non-trivial $E_8$ bundle is thus that over an $S^4$ whose transition functions 
on the $S^3$ equator lie in $\pi_3(E_8)$.  Due to the absence of other relevant homotopy 
classes, $E_8$ bundles over manifolds of dimension $3 < d < 16 $ are topologically 
classified entirely by the transition functions on the $S^3$ equators of $S^4$'s in the 
$4$-skeleton of the base manifold \WittenTools. These are measured by the restriction 
of the first Pontrjagin class $p_1$, which is the exterior derivative of the 
Chern-Simons $3$-form $C_3$ on each coordinate patch \WittenTools, to the given $S^4$.
$E_8$ bundles over $11$-manifolds are thus topologically classified by the specification of
a $3$-form $C_3$, a remarkable fact that depends crucially on the simple low-dimensional
topology of $E_8$.

The basic monopole in such bundles is thus a codimension $5$ object supporting $4$-form flux 
such that the integral of $p_1$ over an $S^4$ linking the defect is the monopole number,
\eqn\sfourflux{
\int_{S^4} \frac{G_4}{2\pi} = n \in \IZ ,
}
where $G_4=dC_3=dTr~(A\wedge F + \frac{2}{3}A\wedge A\wedge A)$.
There is also a codimension $4$ instanton such that the integral of $p_1$ 
over a transverse $4$-plane is non-zero.  Such a bundle can be 
trivialized inside and outside any $3$-sphere in this plane, with the 
transition functions on this linking $S^3$ classified by $\pi_3(E_8)$.  
If we restrict to configurations which are compactly supported 
in the transverse plane, the integral of $p_1$ over the transverse 
$4$-plane is thus an integer counting instanton number.  
Such an instanton can be produced by considering a 
monopole-antimonopole pair whose fluxlines run from one to the other; the integral of $p_1$ over a 
transverse $4$-plane between them is thus quantized, with the choice of orientation specifying 
whether this plane links the monopole or antimonopole and thus fixing the sign.
If the flux takes delta-function support
in the transverse plane, this is a zero-radius instanton Poincare dual to 
the first Pontrjagin class of the bundle.

Due to the magic of $E_8$, 
$$p_2=p_1\wedge p_1=\frac{G_4\wedge G_4}{16\pi^2},$$ 
a relation that would not hold had we considered for example $U(N)$ bundles.
Thus $p_2$ does not reveal any new topology not already contained in $G_4$.  
However, since we can always pull the codimension $5$ defects to infinity, 
$p_2$ can represent a charge in compactly supported cohomology.
For example, consider a bundle such that the integral of $p_2$ over 
some $8$-plane is non-zero; this reveals the presence of a codimension $8$ object 
Poincare dual to $p_2$.
Since we can express $p_2$ as the exterior derivative of a $7$-form $G_7$, 
we can relate this integral over an $8$-plane to an integral over its ``$S^7$ at infinity'' 
(again, we are looking at compactly supported cohomology) to get
$$\int_{\IR^8} p_2 = \int_{S^7} \frac{G_7}{2\pi} = k \in \IZ,$$
so the codimension $8$ objects are quantized and localized.  
There is again an associated codimension $7$ 
``instanton'' (properly, this is an intersection of codimension $4$ instantons) 
such that the integral of $G_7$ over a transverse $7$-plane is non-zero.
Instanton number is quantized in a more subtle way here,
since there is no homotopy class directly counting these instantons.  
However, since these codimension $7$ instantons can be constructed as 
the flux stretching between a codimension $8$ monopole-antimonopole pair, 
a quantization condition applies.

The role of these codimension $7$ and $8$ objects is more transparent when we consider the first
non-trivial AHSS differential for such bundles, 
\eqn\DfourNoTorsion{
d_4 = G_4\cup + {\rm [Torsion]}.
}
Ignoring torsion for the moment, this differential enforces for example the condition
$$d*G_4 =G_4\wedge G_4.$$
This reflects the fact that the $G_7$ whose exterior derivative is $p_2$ really is the
dual of $G_4$.  Physically, this equation requires a codimension $5$ object 
wrapping a $4$-cycle supporting $k$ units of $G_4$ flux to be 
the endpoint of $k$ codimension $8$ objects.

This classification has an immediate reading in terms of the conjecture discussed above.
The codimension $5$ monopole is the $M5$-brane, the codimension $8$ the $M2$-brane,
while the codimension $4$ and $7$ instantons are the $M$-Theory $MF6$ and $MF3$ Fluxbranes 
discussed by Gutperle and Strominger\GutperleSX. 
Moreover, the AHSS differential precisely effects the $11d$ supergravity equation of motion 
$d*G_4=G_4\wedge G_4,$
which implies that an $M5$ wrapping a $4$-cycle supporting $k$ units of $G_4$ flux 
must be the endpoint of $k$ $M2$-branes, a familiar result, and ensures the Dirac quantization
of the $M2$ and $MF3$ branes.

Returning briefly to \DfourNoTorsion, the torsion terms can be studied by
checking when the sign of the Pfaffian of the Dirac operator can be made well defined for
the fermion contribution to a path integral describing an open M2-brane via the inclusion
of some chiral $2$-form.  In particular if the M2-brane wraps a circle we recover the
familiar obstruction $W_3 + H$ from \FreedWitten.
We reserve further discussion of $11d$ torsion until Section $6$; 
about $10d$ torsion we will say more shortly.

At this point it is clear that the soliton spectrum of the various perturbative string theories 
should be reproduced by compactifying the base manifolds of our $11d$ $E_8$ bundles, since it has
precisely reproduced the $M$-Theory solitons from which they descend.
Explicitly studying the dimensional reduction of
the $E_8$ bundle will reveal several interesting details, including an intrinsically
$10d$ classification of IIA solitons treating NSNS and RR solitons largely symmetrically,
to which we now turn.

\newsec{Type IIA and $K$-Theory from $\LE$}

Consider an $E_8$ bundle $F$ over an $11$-manifold $Y$ 
with a circle action that commutes with the transition functions.  
Let $X$ be the $10d$ 
space of orbits of the circle action.  Sections of $F$
thus define sections of an $\LE$ bundle $E\to X$.

Let's pause to review the topology\foot{For a more extensive discussion of such
(possibly centrally extended) loop algebras and the topology of their canonically associated 
group manifolds, see \LoopBook.} of $\LE$.  By the canonical homotopy-lowering map,
$\pi_p(\LE)=\IZ$ for $p=2,14,22,...$, and trivial otherwise.  The low-dimensional 
cohomology is similarly simple,
$$H^{even}(\LE)=\IZ\ \ \ \ \ \ H^{odd}=0.$$
Since $H_2(\LE)=\IZ$, $\LE$ admits a central extension given by a single positive integer.  
This centrally extended Kac-Moody algebra has a canonically associated group manifold, 
both of which we shall denote by $\aE$ in a heinous abuse of notation.  
The topology of $\aE$ differs from that of $\LE$ in several important ways.  In particular, 
$\pi_2(\aE)$ is trivial\foot{The triviality of $\pi_2({\tilde L}G)$ depends 
only on $G$ being simple and simply connected.  
This is essentially the statement that $LG$ admits a single universal central extension 
of which all others are cosets; see \LoopBook\ for an extensive discussion 
of the topology of centrally extended algebras.}, 
and its low-dimensional cohomology is
consequently different from that of $\LE$.

We now return to our $10d$ and $11d$ bundles.
For every soliton or defect in $F$ there is a soliton or defect in $E$.
However, the $10d$ bundle has a generalization which does not lift,
measured by the integer central extension of $\aE$.  Since 
$\pi_3(E_8)=\IZ\neq\pi^*(\pi_2(\aE))$, where $\pi^*$ is the
pullback along the circle fibration projection map,
the central extension of $\aE$ obstructs a lift to $11d$.
Correspondingly, Type IIA string theory has a single integer, 
the $0$-form field strength $G_0$, which is the obstruction to lifting to $M$-Theory.  
Domain walls over which this integer jumps, $D8-$branes, similarly cannot be lifted.  
We thus conjecture that the central extension $k$ of this $\aE$ bundle over $10d$ 
measures the cosmological ``constant'' of (massive) IIA, $G_0$, as
\eqn\GoConj{
G_0 = k.
}
That a lift is indeed possible when $G_0 =0$ fixes a possible additive constant to 
zero\foot{Notice that this proposal is reminiscent to the situation in AdS/CFT, and particularly
$AdS_3\times S^3\times T^4$ in which the cosomological constant on the $AdS_3$ is 
determined by the central charge of the ${\hat {sl}}_2$ 
affine Lie algebra of the boundary WZW model.
We thank Liat Maoz for reminding us of this relationship.}.

The distinct topology of the centrally extended $\aE$ implies that the spectrum
of stable, consistent D-branes is altered in the presence of $D8$-branes.  
In particular, characteristic classes which are torsion when the central extension is non-vanishing
will reveal instabilities of various brane configurations in the presence of $G_0$ which may be stable
in the absence thereof, or vice-versa.  We are thus led to study the complete topology of $\aE$, 
including torsion, which will provide explicit, testable predictions about the (in)stability of brane
configurations in massive IIA\MeWIP. 

Since the homotopy and cohomology groups of $\LE$ agree with those of 
$PU(\infty)=\IC P^\infty=K(\IZ,2)$ up to\foot{$K(\IZ,2)$ is by definition 
the space whose homotopy classes are all trivial 
except for $\pi_2(K(\IZ,2))=\IZ$.  It is realized for example by 
$\IC P^\infty$ which appears in the consideration {\it a la} Sen of 
D-brane classification via non-trivial tachyon bundles associated with the 
gauge bundles over $D$-$\bar D$ pairs.} dimension $14$, 
the classification of RR solitons via $\LE$ bundles differs from that of 
Bouwknegt and Mathai \BoMathai\ only in phenomena related to high (greater than 14) 
dimensional topology\foot{Bouwknegt and Mathai \BoMathai\ argue that IIA D-branes 
are classified by the $K$-Theory of the algebra of sections of a vector bundle 
associated to a $PU(\infty)=K(\IZ,2)$ principal bundle, roughly.}.
Remarkably, the same $\aE$ structure also serves to classify the NS-NS solitons, 
as we now discuss.

	\subsec{NS-NS Solitons from $\LE$}
Since $\pi_2(\LE)=\IZ$, the primary $10d$ $\LE$ defect is codimension $4$, 
i.e. $(5+1)$ dimensional as in $11d$.
An $S^3$ linking $k$ such defects, or more generally any $S^3$ supporting 
$k$ units of $H$-flux as in the SU(2) WZW model, has $\LE$ instanton number 
equal to $k$.  By this we mean that the bundle can be trivialized on the northern 
and southern hemispheres and the transition function is the element $k$ of 
$\pi_2(\LE)$.  The defect is characterized by the fact that, at the 
defect itself, the $\LE$ picture breaks down because the 
circle orbits are not closed.  This $10d$ defect is the reduction
of an $11d$ defect transverse to the $S^1$.  This is precisely the IIA $NS5$-brane
arising from a transverse $M5$-brane.

Similarly, a fundamental IIA string is an $11d$ codimension $8$ soliton whose 
embedding is invariant with respect to the circle action.  
In particular, the $11d$ bundle is then invariant with respect to 
the circle action, so transition functions of the $10d$ bundle consist of 
zero-modes in $\LE$, that is, they inhabit an $E_8$ subgroup.  
In fact the transition functions in $10d$ are just the embedding of those 
in $11d$ into $\LE$, and so the fundamental string is, like the $M2$-brane, 
Poincare dual to the square of the first Pontrjagin class
(the second Pontrjagin class) of this $E_8$ sub-bundle of the $\LE$ bundle.  
This is however not to say that the rest of the $\LE$ is unimportant - 
in particular, the dynamics of the $M2$-brane need not respect the circle 
action, so those of the fundamental string need not restrict themselves to the 
zero mode subgroup at finite coupling.

	\subsec{RR Solitons from $\LE$}

Let's quickly return to the classification of RR solitons via $\LE$ bundles.
The $D4$-brane arises as an $11d$ $5$-defect whose embedding and field configuration 
are invariant under the circle action.  Similarly to the F-string it can be realized
with an $E_8\subset\LE$.  
It is characterized by the fact that each linking $S^4$ has 
$E_8$ instanton number one.  
The $D2$-brane is a $2+1$-soliton transverse to the circle, and is Poincare dual to
$d*G_4$, a $7$-form related to $p_2$ of the $E_8$ bundle by the canonical dimension 
lowering map between characteristic classes of a space and its loop space.
The $D6$-brane arises from a non-trivial circle fibration, 
such that the $\pi_2$ of $\LE$ lifts to the $\pi_3$ of
$E_8$ via a Hopf fibration, while
the $D0$-brane arises as usual as a momentum mode along the $S^1$ fibers.
In both cases the associated flux arises from the KK gauge field, the branes
representing trivial $E_8$ fibrations over the $11$-fold.

Finally, as discussed above, $D8$-brane number is connected to the central extension of $\aE$.  
Thus, while the $D8$-brane does not appear to have a simple geometric interpretation 
in terms of an $11d$ $E_8$ soliton, it has a deep connection to the $\aE$ structure in $10d$.  
This connection may provide insight into the connection between $11d$ gravity 
and the $E_8$ structure\HoravaUday.

	\subsec{Fluxbranes from $\LE$}

The $11d$ $E_8$ origin of IIA Fluxbranes is similarly automatic; its reading in terms
of $\LE$ follows naturally.  
The simplest example is the direct dimensional reduction of the codimension-$4$ $E_8$ fluxtube,
which gives the NS-NS $F6$ in IIA.  
Similarly, a codimension-$4$ fluxtube which wraps the $M$-Theory circle 
remains a codimension-$4$ fluxtube - this is the IIA RR $F5$-brane.
Analogously, the codimension-$8$ fluxtube reduces transversely to the RR $F3$-brane and, 
wrapping the $M$-Theory circle, to the NSNS $F2$-brane.  The $F1$ and $F7$ arise as fluxtubes
associated to the nontrivial bundles of the $D0$ and $D6$ branes, respectively.
Thus we realize the full spectrum of RR and NSNS $Fp$-Branes discussed by 
Gutperle and Strominger \GutperleSX\ in terms of $\LE$, as expected.

	\subsec{$K$-Theory from $\LE$ and Indiscretions regarding Torsion}
We have seen how the classification of both NSNS and RR solitons in Type IIA arises from the
classification of $\LE$ bundles in $10d$, these derived from a fundamental $E_8$ structure
in $M$-Theory.  Due to the remarkable topology of $\LE$, this reproduces the
conjectured $K$-Theoretic classification for RR charges and fields.  We would now like to
connect this construction with the AHSS approximation to the $K$-Theoretic classification.
In the remainder of this section we will use the language of $M$-branes and D-branes for
simplicity and clarity; in light of the above discussion, it should be clear that the entire
discussion can be phrased explicitly in terms of $11d$ $E_8$ bundle information.

The classifying group of solitons in $M$-Theory is a refinement of cohomology 
obtained by taking the quotient with respect to a series of differentials that reflect the 
fact that some configurations are anomalous and so should not be included, while others
are related by dynamical processes and so must carry the same conserved charges(see eg \MMS).  
For example, an $M5$-brane wrapping a $4$-cycle that supports $k$ units of $G_4$ flux leads 
to an anomaly that, neglecting torsion, can be canceled if $k$ $M2$-branes end on the 
$M5$.  Thus some $M5$-brane wrappings are anomalous and some $M2$-brane configurations 
(such as $k$ $M2$'s and the vacuum) are equivalent, this following from
the $11d$ supergravity equation of motion 
$$d*G_4=G_4\wedge G_4.$$
The left hand 
side of this equation is the intersection number of $M2$-branes with a sphere linking 
the $M5$, and the right is roughly the integral of the $G_4$ flux over the 4-cycle 
wrapped by the 5-brane.  Both of these numbers are measured in units of the 8-form 
Poincare dual to the $M2$-branes.  In the absence of $M2$-branes ending on the $M5$'s, 
this supergravity constraint is summarized\foot{This was seen in type II in \MMS.} 
by requiring that the following ``differential'' annihilate the $G_4$ flux 
$$d_4 G_4 =G_4\wedge G_4 + {\rm [Torsion]}.$$  
We expect that the torsion terms 
are nontrivial because, for example, $G_4$ is half-integral when the $M5$ brane wraps 
a 4-cycle with non-vanishing $w_4$ \WittenFiveBrane.  Also, as we will soon see, its dimensional 
reduction is nontrivial.

The classification for IIA follows from dimensional reduction of this $M$-Theory story.
There are three distinct classes of reductions of this constraint to IIA, 
reflecting three possible locations
of the $M$-Theory circle $x^{11}$ in the above scenario.  First consider an 
$M5$-brane wrapping $x^{11}$ which is not in the $4$-cycle, so that the anomaly-canceling 
$M2$-branes do wrap $x^{11}$.  This leads to an anomaly condition requiring 
$F$-string insertions on a $D4$ as follows.  The $M5$-brane wraps $x^{11}$ and so the $G_4$ flux that 
it generates has no $11$ component; it is thus not Kaluza-Klein reduced.  Similarly, the $4$-cycle 
does not wrap and so the $G_4$ supported on the $4$-cycle is not reduced.  Thus the $10d$ 
anomaly condition arising from this situation is identical to the $11d$ condition: 
$$d_4 G_4=G_4\wedge G_4 + {\rm [Torsion]},$$ 
now a $10d$ constraint with $G_4$ identified with the $4$-form RR fieldstrength.  

Next consider the case in which both the $M5$-brane and the $4$-cycle wrap $x^{11}$, 
yielding a $D4$ with $D2$ insertions as follows.  The $G_4$ 
flux sourced by the $M5$-brane is still not reduced, but now the $4$-cycle is reduced to a 
$3$-cycle, the $G_4$ flux it supports dimensionally reduced to the 
$3$-form $H$.  The resulting anomaly constraint is thus
$$d_3 G_4=H\wedge G_4 + {\rm [Torsion]}.$$  
This is a well-known differential from the AHSS 
for twisted $K$-Theory \Rosenberg, which was seen to be the relevant 
constraint in \MMS.  
In particular the torsion correction was seen to be $Sq^3 G_4$.  

The final case involves an $M5$-brane not wrapping $x^{11}$, reducing to an 
$NS5$-brane with $D2$-brane insertions.  In this case the $4$-form flux is dimensionally 
reduced to $H$ while the flux in the $4$-cycle is not reduced, yielding the constraint 
$$d_4 H=G_4\wedge H + {\rm [Torsion]}.$$  
The torsion in this case is as yet poorly understood.

Combining these three constraints, as well as the AHSS conditions on other RR fluxes, we hope to 
arrive at a $K$-Theoretic classification of both NSNS and RR charged objects in IIA.  We expect 
this classification to be T-dual to the S-duality covariant classification in \SquareThree.
Independently of our proposal, it would be interesting to better understand the $11d$ lifts 
of the other constraints on RR fluxes.

For example, anomaly cancellation on a $D2$-brane in IIA wrapping a $3$-cycle $C$ with $k$ units 
of $H$ flux requires $k$ $D0$-brane insertions.  Lifting this to $M$-Theory we learn that, while 
we know of no restrictions on what cycle an $M2$-brane may wrap, if it wraps a $3$-cycle $C$ such 
that 
$$\int_{C\times S^1} \frac{G_4}{2\pi} =k\neq 0$$ 
then $k$ units of momentum around $x^{11}$ must be absorbed by the brane.  
To get an intuitive understanding of the physics at work\foot{See also the beautiful discussion 
in \TaylorZA, which addresses an analogous effect for dielectric branes 
in a non-compact geometry.}, let us pretend that $C$ is 
a $2$-cycle times the time direction, with a constant $H$ flux density, and then 
KK reduce on the $2$-cycle.  Before reducing, this
corresponds to a constant flux of $D0$-branes incident on the $D2$-brane in IIA,
while in $M$-Theory this corresponds to a steady injection of $p^{11}$ into the $M2$.
KK reducing, the $G_4$-flux reduces to an electric field along the circle, while the
$M2$-brane reduces to a particle charged under this field. 
This flux drives the charged particle to accelerate around the 
circle with a constant acceleration, that is, to absorb $p^{11}$ at a constant rate.
The anomaly condition lifted to $M$-Theory is thus simply 
$F=ma$!
Although we do not understand the deep connection of the $M$-Theory $E_8$ bundle 
to gravity, this relation between $G_4$ and $p^{11}$ is perhaps a significant clue.

\newsec{The Heterotic String and the Small Instanton Transition}

Consider now an $E_8$ bundle over an $11$-fold $X=M\times S^1/\IZ_2$.  
The bulk bundle naturally restricts to two $10d$ $E_8$ bundles, 
one over each of the two boundary components.  
At this point the realization of the various objects in Heterotic 
string theory in terms of instantons of the $E_8$ bundle follows 
naturally from the beautiful arguments of \HW.
For example, an $M2$-brane stretching between the two boundary components is precisely the
strongly-coupled fundamental Heterotic string.  Moreover, anomaly considerations descend naturally.
In 11-d, there is a mod 2 relation between the Pontrjagin classes of the $E_8$ bundle, $w(F\to Y)$,
and that of the base manifold's tangent bundle, $w(TY)$ - thus for example 
$G_4 = w_4(TY)/2$\ . 
This condition reduces on the induced bundle over the orbifold fixed point to the $10d$ condition,
which arises from a gravitational anomaly \HW\HoravaQA.

It is easy to see the Heterotic $5$-brane arising from the bulk $E_8$ bundle.
Recall that the $11d$ $E_8$ $5$-defect is defined such that a
$4$-sphere linking the $5$-defect has instanton number one.
Consider a parallel $11d$ $5$-${\bar 5}$ pair separated a finite distance 
in a transverse direction, $y$, of $\IR^{(10,1)}$.  
For every point $y_p$ there is a $10d$ bundle given by the restriction of 
the $11d$ bundle to the $10d$ slice $y=y_p$.  Since any $4$-plane in 
the slice $y=y_p$, with $y_p$ between the two defects, 
links one or the other of the defects,
the $10d$ bundles over points between the two $5$-defects have instanton number 
$\pm 1$, the sign fixed by choice of orientation, while the $10d$ bundles over 
points not between the two defects have instanton number zero.  Since the
$10d$ bundles over points between the $11d$ defects are non-singular, their
instantons are ``large''.  The singular $10d$ bundles which contain the $11d$ $5$-defects, 
by contrast, contain ``small'' instantons.  These are the Heterotic $5$-branes.

Next consider a similar configuration with the two defects pulled away to infinity, 
leaving a single codimension-$4$ instanton stretched along the coordinate $y$ and 
taking compact support in the transverse $4$-plane.  If we pinch the instanton 
over a point $y=y_*$, we can nucleate a $5-{\bar 5}$ pair at $y_*$ and move them away to infinity,
leaving behind no flux in the interval between them.  From the point of view of the $11d$ bundle, 
this is a completely continuous process respecting all conserved charges and symmetries.
From the point of view of the induced $10d$ bundle over any point $y=y_o\ne y_*$, however, 
things look rather odd; the originally large and fluffy instanton shrinks to a singular 
``small'' instanton and then disappears altogether!  

Now consider an $E_8$ bundle over the $11$-manifold $Y=\IR^{(9,1)}\times (\IR/\IZ_2)$, where the
$y$ coordinate along which the $11d$ instanton is extended has been orbifolded by a $Z_2$ reflection.  
If we repeat the pinching-transition over the point $y=0$, which from the point of view 
of the covering space is completely continuous and respects all conservation laws,
as well as the orbifold symmetry, we find a transition in the orbifold theory in which 
a ``large'' instanton in the boundary bundle shrinks to a singular ``small'' instanton 
before disappearing from the boundary and moving into the bulk as an $11d$ $5$-defect, 
i.e. an M$5$-brane.  This is precisely the Heterotic small instanton transition studied 
near one boundary component, as read by the $11d$ $E_8$ bundle.  Note that, 
while the number of boundary instantons $n_{\partial Y}$ is not conserved, 
$n_{\partial Y} + n_{Y}$ is.

\newsec{Speculations about $E_8$ Bundles and $11d$ SUSY}

Since objects to which the $E_8$ gauge connection couples 
in $M$-Theory and string theory
violate at least half of the $32$ $11d$ supercharges,
we should perhaps expect to see gauge bundle information only in situations with 
reduced supersymmetry.  It is thus reasonable to wonder if
the gauge connections inhabit representations 
of only a sub-algebra of the $11d$ superalgebra, representations that in particular 
contain neither gravitons nor gravitinos.  The Chern-Simons 3-form of this connection
can then be set equal to the 3-form in the $11d$ supermultiplet, for example via a 
Lagrange multiplier\foot{We particularly thank Eva Silverstein 
for discussions on this topic.},
$$\delta S \sim  \int_{M^{11}} \alpha (C^M_3 -C^{E_8}_3).$$  
It is worth keeping in mind that both the $M$-Theory $3$-form and the $E_8$ Chern-Simons form respect
an abelian gauge symmetry, since for example under a local $E_8$ gauge transformation with gauge 
parameter $\Lambda$ the CS-$3$-form transforms as $C\to C+d{\rm Tr}(\Lambda F)$, so this action 
is in fact gauge invariant and respects all the requisite symmetries.

Of course, not all bundles in the same topological equivalence class correspond to 
BPS solitons.  Rather, the bundles in each equivalence class are related by 
a change in boundary conditions which does not change the topology; 
in the associated SUGRA class, this corresponds 
(roughly, as the equations of motion are non-linear)
to a shift by a solution to the vacuum equations of motion.  
However, since the topological classification of bundles 
is precisely the classification by charge
(at least up to torsion terms), 
there is some choice of background fields
which does not affect the topological class and yields precisely the BPS soliton.
In particular we attribute an array of classical moduli, such as the size of Heterotic 
instantons, to precisely such a freedom of choice of boundary conditions.

\newsec{Conclusions and Open Questions}

We have argued that the topological classification of $E_8$ bundles in $11d$,
which naturally reproduces the soliton spectrum of $M$-Theory, 
reproduces when reduced on $S^1/\IZ_2$ the spectrum of Heterotic $\ee$,
while reduced on $S^1$ reproduces the spectrum and K-classification 
of RR and NSNS solitons in Type IIA\foot{While we of course do not have a candidate for what the
complete $K$-Theory of $\aE$ bundles is, it should be identical to that of the universal classifying 
group $K(\IZ,2)$ up to corrections involving topology well above $11d$, as discussed 
above.  One might for example attempt to generalize Rosenberg's $K$-Theory, \Rosenberg.}.  
Remarkably, while there appears to be no simple dynamical role for 
$E_8$ in Type IIA, there does appear to be a deep role for its loop group
$\LE$ in the $K$-Theoretic classification of IIA solitons, including in an important way 
its central extension.  The relevance of $E_8$ bundles even for perturbative string theories 
with no dynamical gauge bosons suggests an important role for $E_8$ in 
the construction of the fundamental degrees of freedom of $M$-Theory.

The most obvious open question is how, precisely, an $11d$ gauge theory fits with $11d$ supersymmetry.
This is extremely confusing.  Perhaps a natural place to look for hints to this puzzle is in 
Heterotic $\ee$, where the gauge boson couples in an intricate but natural and beautiful way.
Extending this story to $11d$ would be an exciting advance.

Another obvious omission in our presentation is the absence of 
torsion terms in \DfourNoTorsion.
That this is an important omission is clear from any geometry where, for example, 
an $M5$-brane lies inside not an $S^4$ but some orbifold thereof.
Following \DiaMoWitten, one thus expects
the torsion terms to include some $\IZ$ lift of $sq^4$; 
however, as there is no canonical lift of the $\IZ_2$ Steenrod squares of even rank, 
identifying the correct ``derivation'' is somewhat delicate.  
In the language of Witten, and in the orientable case, 
one might expect the fourth AHSS differential to take the form 
$d_4=\lambda +G_4\cup$.  
However, the sign in front of $\lambda$ is not obvious.  
It could of course be fixed by comparison with the $5$-brane anomaly, 
but would still leave ambiguous the correct torsion terms in non-orientable cases, 
where some lift of the $\IZ_2$ Steenrod square $sq^4$ must obtain.  

One avenue of approach might be 
to identify a canonical lift for the special case of 
$11$-folds with compatible circle actions.  As a first guess, define 
$${\tilde Sq}^4=\pi^*(Sq_3),$$
where $\pi^*$ is the pullback of the projection of the $S^1$ fibration.
From various Adem relations one can argue that this restricts correctly to $sq^4$ if 
$\pi^*(\beta)=sq^2$.
A case where one might test this possibility would be an $M5$-brane wrapping 
$SU(3)/SO(3)\equiv M_5$, whose anomaly requires an $M2$-brane to end upon the 
$M5$-brane. Reducing on an $S^1$ to a $D4$-$D2$, the anomaly arises from 
$Sq^3$ in the $D4$-brane worldvolume, which is canceled by the incident $D2$.  
Pulling back along the $S^1$ fibration, 
$Sq^3$ should lift to a $\IZ$-graded rank-four differential which measures
the correct $10d$ anomaly under bundle projection.
It would be interesting to explicitly check when, 
if ever, such a non-trivial pullback exists, and when it does 
whether it restricts to the $\IZ_2$-graded $sq^4$.  
We leave such questions to future work.


Finally, it would be particularly interesting to revisit the beautiful and delicate calculations
of Diaconescu, Moore and Witten in \DiaMoWitten, who showed that the cancellation of anomalies in IIA
and $M$-Theory agree, though the structures underlying the calculations in the two cases were
apparently unrelated.  DMW read this unlikely agreement as strong 
evidence for the conjecture that RR fields and charges in IIA are indeed classified by $K$-Theory.  
We expect that the IIA computation will take a natural form in terms of 
${\hat E}_8$ bundles, and that in this language the relation to anomaly 
cancellation in $M$-Theory will be immediate.  This would be interesting to check directly.

\bigskip
\bigskip
\bigskip
\bigskip
\noindent{\bf Acknowledgments}

We would like to thank Michal Fabinger, Petr Ho{\v r}ava, Albion Lawrence, John McGreevy, 
Hisham Sati, Mike Schulz, Eric Sharpe, Eva Silverstein and Uday Varadarajan for discussions.  
We are particularly indebted to Petr Ho{\v r}ava for essential insights and motivation early on.
J.~E. particularly thanks Uday Varadarajan for many conversations.
A.~A. especially thanks Eva Silverstein and John McGreevy for many stimulating 
conversations regarding early drafts, and Petr Horava for very enjoyable conversations
on this and related topics.
We thank Don Marolf and Andreas Gomberoff for hospitality at PASI 2002, 
where much of this work was conducted.
The work of A.~A. was supported in part by DOE contract DE-AC03-76SF00515 and 
by an NSF Graduate Fellowship.

\listrefs

\end